\documentclass[aps,prl,twocolumn,floatfix,superscriptaddress,showpacs,amsfonts,amssymb,amsmath,preprintnumbers]{revtex4}
\usepackage{bm}
\usepackage{psfig}
\usepackage{dcolumn}

\newcommand{\z}[1]{{\tt z#1}}
\newcommand{\zABI}[1]{\textcolor{red}{ABI #1}}
\renewcommand{\z}[1]{}
\renewcommand{\zABI}[1]{}

\newcommand{\eq}[1]{Eq.~(\ref{#1})}

\newcommand{\figref}[1]{Fig.~\ref{#1}}

\newcommand{\bea}{\begin{eqnarray}}
\newcommand{\eea}{\end{eqnarray}}
\newcommand{\bal}{\begin{aligned}}
\newcommand{\eal}{\end{aligned}}
\newcommand{\bga}{\begin{gathered}}
\newcommand{\ega}{\end{gathered}}

\newcommand{\bi}[1]{{\bf #1}}

\newcommand{\lt}{\left}
\newcommand{\rt}{\right}
\newcommand{\la}{\langle}
\newcommand{\ra}{\rangle}

\newcommand{\const}{{\rm const}}
\newcommand{\eps}{\varepsilon}

\newcommand{\dd}{\partial}
\newcommand{\vdel}{\bi{\nabla}}
\newcommand{\vu}{\bi{u}}
\newcommand{\vf}{\bi{f}}
\newcommand{\vB}{\bi{B}}

\newcommand{\vk}{\bi{k}}

\newcommand{\lf}{L}
\newcommand{\lvisc}{l_\nu}
\newcommand{\lres}{l_\eta}

\newcommand{\nueff}{\nu_{\rm eff}}

\newcommand{\urms}{u_{\rm rms}}
\newcommand{\usq}{\la u^2\ra}
\newcommand{\Bsq}{\la B^2\ra}

\newcommand{\Pm}{{\rm Pm}} 
 
\renewcommand{\Re}{{\rm Re}} 
 
\newcommand{\Rm}{{\rm Rm}} 
\newcommand{\Rmc}{\Rm_{\rm c}} 
\newcommand{\Rmcmax}{\Rm_{\rm c}^{\rm(max)}} 
\newcommand{\Rmcinf}{\Rm_{\rm c}^{(\infty)}} 

\newcommand{\gammainf}{\gamma_\infty}

\begin{document}

\preprint{{\em Phys.\ Rev.\ Lett.}~{\bf 98}, 208501 (2007); E-print {\tt astro-ph/0702291}}

\title{Numerical demonstration of fluctuation dynamo at low magnetic Prandtl numbers} 
\author{A.\ B.\ Iskakov}
\affiliation{Department of Physics and Astronomy, UCLA, Los Angeles, California 90095-1547, USA}
\author{A.\ A.\ Schekochihin}
\altaffiliation{Author to whom correspondence should be addressed; 
electronic mail a.schekochihin@imperial.ac.uk}
\affiliation{Plasma Physics Group, Blackett Laboratory, Imperial College, London SW7~2BW, United Kingdom}
\affiliation{King's College, Cambridge CB2 1ST, United Kingdom}
\affiliation{DAMTP, University of Cambridge, Cambridge CB3 0WA, United Kingdom}
\author{S.\ C.\ Cowley}
\affiliation{Department of Physics and Astronomy, UCLA, Los Angeles, California 90095-1547, USA}
\affiliation{Plasma Physics Group, Blackett Laboratory, Imperial College, London SW7~2BW, United Kingdom}
\author{J.\ C.\ McWilliams}
\affiliation{Department of Atmospheric Sciences, UCLA, Los Angeles, California 90095-1565, USA}
\author{M.\ R.\ E.\ Proctor}
\affiliation{DAMTP, University of Cambridge, Cambridge CB3 0WA, United Kingdom}

\begin{abstract}
Direct numerical simulations of incompressible nonhelical 
randomly forced MHD turbulence are used to demonstrate for the first time 
that the fluctuation dynamo exists in the limit of large magnetic Reynolds 
number $\Rm\gg1$ and small magnetic Prandtl number $\Pm\ll1$. 
The dependence of the critical $\Rmc$ for dynamo on the 
hydrodynamic Reynolds number $\Re$ is obtained for 
$1\lesssim\Re\lesssim6700$. In the limit $\Pm\ll1$, 
$\Rmc$ is about three times larger 
than for the previously well established 
dynamo at large and moderate Prandtl numbers: 
$\Rmc\lesssim 200$ for $\Re\gtrsim6000$ 
compared to $\Rmc\sim60$ for $\Pm\ge1$.
Is is not as yet possible to determine numerically 
whether the growth rate of the magnetic energy 
is $\propto\Rm^{1/2}$ in the limit $\Rm\to\infty$,
as it should if the dynamo is driven 
by the inertial-range motions at the resistive scale. 
\end{abstract}


\maketitle

\paragraph{Introduction.} 
The amplification of magnetic field by 
turbulent fluid motion, or {\em dynamo}, 
is believed to be the cause of cosmic magnetism 
\cite{Roberts_Glatzmaier,Ossendrijver_review,Widrow}. 
Two types of turbulent dynamo should be distinguished. 
The first is the {\em mean-field dynamo} defined as the growth of 
magnetic field at scales larger than the outer (energy-containing) 
scale $\lf$ of the turbulent fluid motion. 
The second, which is the focus of this Letter, 
is the {\em fluctuation dynamo} (or small-scale dynamo) 
defined as the growth of magnetic-fluctuation energy 
at or below the outer scale \footnote{In the presence of 
a large-scale field, small-scale magnetic 
fluctuations can also be generated by turbulent tangling, or 
induction, --- a mechanism physically distinct from the 
fluctuation dynamo. See A.~A.~Schekochihin {\em et al.}, 
New J.\ Phys. (to be published), and references therein.}. 

Reflexively, one tends to think of turbulence as 
an effective mixing mechanism rather than a constructive agent. 
It is then a remarkable idea that random motions can amplify a magnetic field. 
The currently accepted qualitative explanation of how the fluctuation 
dynamo is possible is based on the notion of the random Lagrangian stretching 
of the field by the fluid motion 
\cite{Batchelor,Moffatt_Saffman,Zeldovich_etal,SCTMM_stokes}. 
This picture depends on the assumption that the scale of the stretching 
(which is the viscous scale $\lvisc$, because the motions there have the 
largest turnover rate) 
is larger than the scale of the field that is stretched (the resistive scale $\lres$). 
Whether this is true depends on the magnetic Prandtl number, 
$\Pm=\Rm/\Re$, which, in most natural systems, is either very large or very small. 
When $\Pm\gg1$, we have indeed $\lres/\lvisc\sim\Pm^{-1/2}\ll1$ 
(see \cite{SCTMM_stokes} and references therein). 
In contrast, when $\Pm\ll1$ (while both $\Re\gg1$ and $\Rm\gg1$), 
one expects $\lres/\lvisc\sim\Pm^{-3/4}\gg1$ 
\cite{Moffatt}, so the resistive scale is in the middle 
of the inertial range, asymptotically far away both from the viscous 
and outer scales. 
Can the field still grow? 
In the absence of a mechanistic model of the field amplification, 
the problem is quantitative: the stretching and turbulent 
diffusion are of the same order at each scale in 
the inertial range, so which of them wins is not obvious. 
For over 50 years, 
the resolution of this problem has fascinated and confounded several 
generations of scholars, who were intermittently convinced that 
the fluctuation dynamo did or did not exist. 

With the advent of modern scientific computing, 
numerical simulations have been used to build a 
case for dynamo as a very generic property of random and chaotic flows 
\cite{Meneguzzi_Frisch_Pouquet,Cattaneo,Brandenburg,SCTMM_stokes,Ponty_etal1}. 
This case has recently been strengthened by a successful laboratory 
dynamo in a geometrically unconstrained 
turbulence of liquid sodium \cite{Monchaux_etal}. 
However, both in the computer and in the laboratory, 
it is nearly impossible to access the values of $\Re$ and $\Rm$ 
that are sufficiently large to resemble real astrophysical situations. 
It has been especially difficult to model the case of 
$\Re\gg\Rm\gg1$, corresponding to the limit of low $\Pm$. 
The computational challenge in this limit is to resolve two scale 
separations: $L\gg\lres\gg\lvisc$. 
The low-$\Pm$ limit is encountered in 
the liquid-metal cores of planets ($\Pm\sim10^{-5}$),  
the solar convective zone ($\Pm\sim10^{-7}-10^{-4}$), 
protostellar disks, etc. 
In the absence of a proof of the fluctuation dynamo at low $\Pm$, 
the case for the dynamo origin of the small-scale magnetic fields 
in such systems (e.g., the observed fields in the solar photosphere)
has been based on simulations done in 
the opposite regime of $\Pm\ge1$ \cite{Cattaneo}. 

Previous numerical investigations of the onset of the 
fluctuation dynamo at low $\Pm$ 
\cite{SCMM_lowPm,Haugen_Brandenburg_Dobler,SHBCMM_lowPm2} 
revealed that the critical magnetic Reynolds number 
$\Rm_c$ for dynamo increased with $\Re$. 
In this Letter, we report that 
$\Rmc$ eventually reaches a finite limit as $\Re\to\infty$, 
i.e., {\em we demonstrate for the first time that the fluctuation dynamo at 
asymptotically low $\Pm$ exists.} The most important outstanding issue 
is whether the dynamo we have found is driven by the 
inertial-range motions at the resistive scale --- if it is, 
its growth rate should be proportional to $\Rm^{1/2}$, which would make 
it a dominant (and universal!) field-amplification effect compared 
to any mean-field dynamo due to the outer-scale motions. 

\begin{figure}[t]
\centerline{\psfig{file=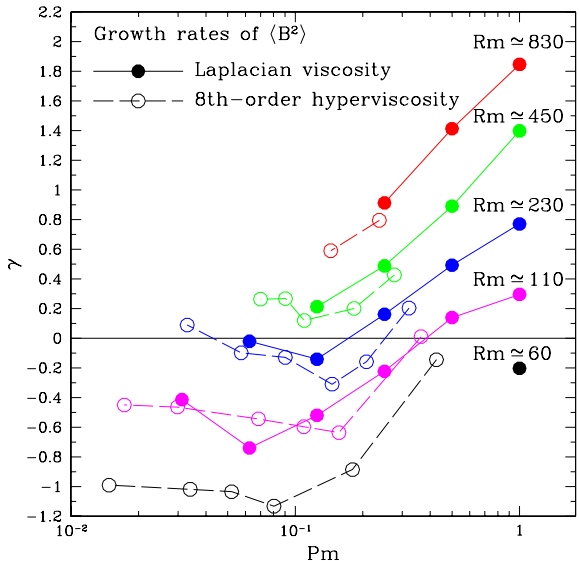,width=8cm}}
\caption{\label{fig_gamma} Growth/decay rate of $\Bsq$ vs.~$\Pm$ for 
$\eta=4\times10^{-3}$ ($\Rm\sim60$), 
$\eta=2\times10^{-3}$ ($\Rm\sim110$),
$\eta=10^{-3}$ ($\Rm\sim230$),
$\eta=5\times10^{-4}$ ($\Rm\sim450$), 
and $\eta=2.5\times10^{-4}$ ($\Rm\sim830$).} 
\vskip-0.3cm
\end{figure}

\paragraph{Numerical Set Up.} 
The equations of incompressible magnetohydrodynamics 
\bea
\label{u_eq}
&&\dd_t\vu + \vu\cdot\vdel\vu = -\vdel p - \nu_n|\vdel|^n\vu
+ \vB\cdot\vdel\vB + \vf,\\
\label{B_eq}
&&\dd_t\vB + \vu\cdot\vdel\vB = 
\vB\cdot\vdel\vu + \eta\nabla^2\vB
\eea
are solved in a periodic code of size 1, using a standard pseudospectral 
code. Here $\vu$ is the velocity and $\vB$ the magnetic field (in velocity 
units). The density is constant and equal to~1. 
The velocity is forced by a random {\em nonhelical} 
homogeneous white-noise body force, which injects energy 
into the velocity components with wave numbers $|\vk|\le\sqrt{2}\, k_0$, 
where $k_0=2\pi$ is the box wave number. 
Because the forcing is a white noise, 
the average injected power is fixed: $\eps=\la\vu\cdot\vf\ra=1$, 
where the angle brackets stand for volume and time averaging.
The numerical integration is continued only for as long as is needed 
to obtain a converged value of the growth/decay rate. 
In all of our runs, the magnetic field is energetically much weaker 
than the velocity at all times, so the Lorentz force is never important.

The maximum resolution we could afford was $512^3$. 
In order to increase the range of accessible Reynolds numbers, 
we performed simulations both with the Laplacian viscosity 
($n=2$ in \eq{u_eq}) and with the 8th-order hyperviscosity ($n=8$). 
We define 
$\Rm = {\usq^{1/2}/\eta k_0}$,
$\Re = {\usq^{1/2}/\nu k_0}$,
$\Pm = {\nu/\eta}$, 
where $\nu=\nu_2$ for the Laplacian runs and $\nu=\nueff={\eps/\la|\vdel\vu|^2\ra}$ 
for the hyperviscous ones. We believe that using hyperviscosity is justified 
for $\Pm\ll1$ because the resistive scale in this limit is 
much larger than the viscous scale, $\lres\gg\lvisc$, and 
the magnetic properties of the system should be independent of the 
form of viscous regularization. 

\paragraph{Existence of the Dynamo.} 
In \figref{fig_gamma}, we show the growth/decay rates 
of the magnetic energy $\Bsq$ vs.\ $\Pm$ 
for five sequences of runs, each with a fixed value of $\eta$. 
Thus, decreasing $\Pm$ is achieved 
by increasing $\Re$ while keeping $\Rm$ fixed. 
The growth rates are 
calculated via a least-squares fit to the evolution 
of $\ln\lt(\Bsq\rt)$ vs.~time. 
We find that 
as $\Pm$ is decreased, the growth rate decreases, 
passes through a minimum and then saturates a constant 
value, i.e., at fixed $\Rm$,
$\gamma(\Rm,\Re)\to\gammainf(\Rm)=\const$ as $\Re\to\infty$. 
It is natural that such a limit exists because 
$\lres\gg\lvisc\sim\lf\Re^{-3/4}\to0$ 
and one cannot expect the magnetic field to ``know'' 
exactly how small the viscous scale is. 
The nontrivial result is that, as $\Rm$ increases, 
the entire curve $\gamma(\Re,\Rm)$ is lifted upwards, 
so the asymptotic values $\gammainf(\Rm)>0$. 
Although we cannot at current resolutions determine these 
positive asymptotic 
values, our judgement is that \figref{fig_gamma} provides sufficient 
evidence for claiming that such values exist and are positive. 

\begin{figure}[b]
\vskip-0.3cm
\centerline{\psfig{file=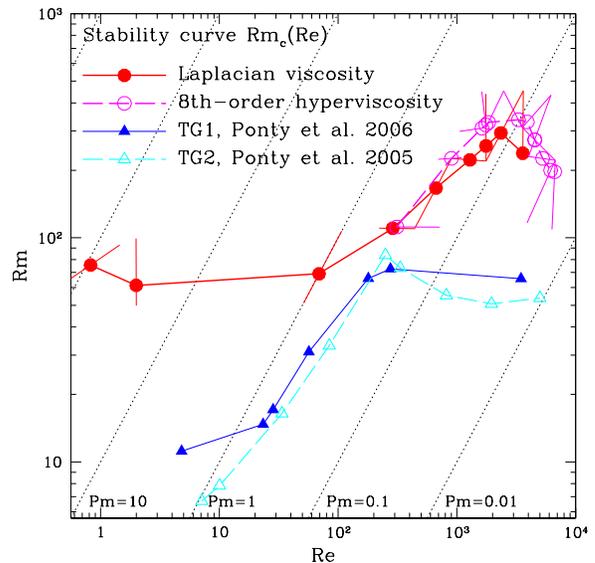,width=8cm}}
\caption{\label{fig_Rmc} The stability curve $\Rmc$ vs.\ $\Re$. 
``Error bars'' connect $(\Re,\Rm)$ for decaying and growing runs 
used to obtain points on the stability curve. Stability curves 
based on the Laplacian and hyperviscous runs are shown 
separately. For comparison, we also plot the $\Rmc(\Re)$ curve 
obtained in simulations employing TG1 \cite{Ponty_etal2}
and TG2 forcing \cite{Ponty_etal1} (the three highest-$\Re$ points 
in the latter case were obtained by large-eddy simulations).
The values of $\Re$ and $\Rm$ are recalculated according to our 
definitions, using the forcing wavenumber $k_0$, rather than 
the dynamical integral scale as in \cite{Ponty_etal1,Ponty_etal2}.} 
\end{figure}

\begin{figure*}
\begin{tabular}{ccc}
\psfig{file=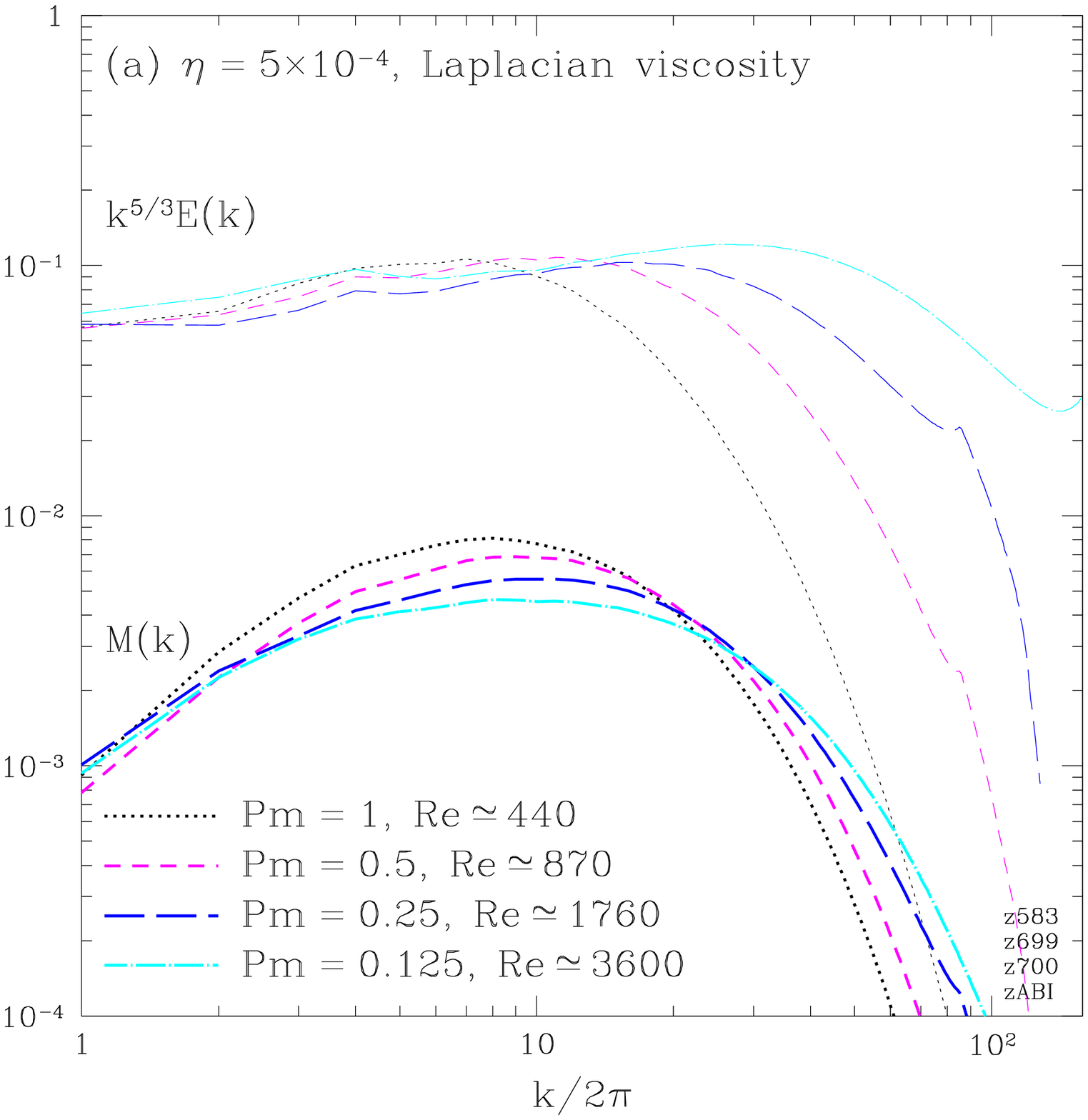,width=8cm} && \psfig{file=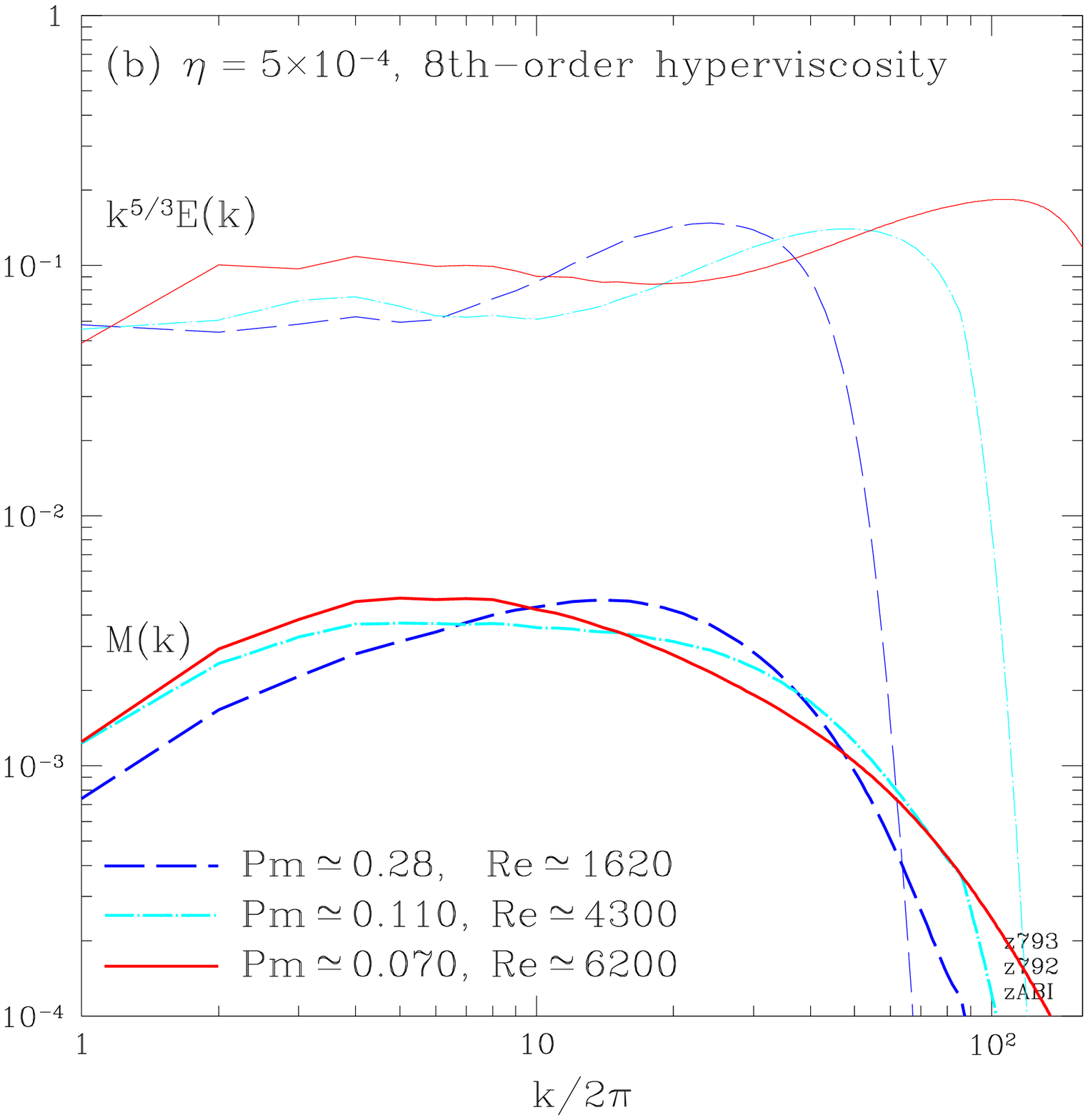,width=8cm}
\end{tabular}
\caption{\label{fig_spRm} Normalized spectra of kinetic energy (compensated by $k^{5/3}$) 
and (growing) magnetic energy for fixed $\eta=5\times10^{-4}$ 
(the $\Rm\sim450$ sequence from \figref{fig_gamma}) and increasing $\Re$: 
(a) Laplacian runs, 
(b) hyperviscous runs.}
\vskip-0.3cm
\end{figure*}

\paragraph{The Stability Curve.} In \figref{fig_Rmc}, we show the  
reconstructed stability curve $\Rmc(\Re)$: 
each point on the curve is obtained by linear interpolation 
between a decaying and a growing case. 
We see that $\Rmc$ increases with $\Re$, 
reaches a maximum $\Rmcmax\sim350$ at $\Re\sim3000$, 
and then decreases. 
Since the exact value of the viscous scale cannot matter 
when $\Re\gg\Rm$, 
$\Rmc(\Re)\to\Rmcinf=\const$ as $\Re\to\infty$.
We cannot as yet obtain this asymptotic value. 
Unless the stability curve has multiple extrema 
(which we consider unlikely), $\Rmcinf\lesssim200$. 
Note that this value is only about 3 times larger 
than the well-known critical value $\Rmc\simeq60$ 
for the fluctuation dynamo at $\Pm\ge1$ 
\cite{Meneguzzi_Frisch_Pouquet,SCTMM_stokes,Haugen_Brandenburg_Dobler}. 

Considering that there is a small but measurable difference 
between the stability curves for the Laplacian and hyperviscous 
simulations, how universal are our results? 
We believe it is plausible to argue, as we did above, 
that the existence of the dynamo in the limit 
$\Rm\ll\Re\to\infty$ does not depend on the nature of 
the viscous regularization. We also think 
that the asymptotic value $\Rmcinf$ is likely to be robust. 
However, the full functional dependence $\Rmc(\Re)$ 
is certainly not universal. Indeed, let us consider 
what determines the shape of the stability curve 
in the transition region between the high- and low-$\Pm$ 
limits. When $\Re<\Rm$, $\lvisc>\lres$. 
As $\Re$ is increased, the spectral bottleneck associated with 
the viscous cutoff moves past the resistive scale $\lres$ 
until finally $\lvisc\ll\lres$ and 
the resistive scale is in the inertial range. 
This transition is illustrated by \figref{fig_spRm}. 
The properties of the velocity field around the 
viscous scale are obviously not independent of the 
viscous regularization. 
Therefore, the $\gamma(\Re,\Rm)$ and $\Rmc(\Re)$ 
cannot be universal in the transition region. 
In particular, since the bottleneck is narrower 
for the hyperviscous runs, 
the transition in the parameter space should be sharper. 

\paragraph{Magnetic-Energy Spectrum.} 
The shape of the magnetic-energy spectrum 
is qualitatively different for $\Pm\ge1$ 
and $\Pm\ll1$ (see \figref{fig_spRm}). 
At $\Pm$ above and just below 
unity, the spectrum has a positive slope and its peak is at the 
resistive scale. This is typical for the fluctuation 
dynamo at $\Pm\ge1$ --- in the limit $\Pm\gg1$, the 
Kazantsev \cite{Kazantsev} $k^{+3/2}$ spectrum emerges \cite{SCTMM_stokes}.
As $\Pm$ is decreased, 
the spectrum flattens and then appears to develop a negative slope
in the inertial range. 
At current resolutions, it is not possible to determine 
definitively what the asymptotic spectral slope is 
and whether the spectral peak is independent of $\Rm$ 
or moves with the resistive scale as $k_{\rm peak}\propto\Rm^{3/4}$. 

\paragraph{Comparison with Simulations with a Mean Flow.} 
Several authors 
\cite{Ponty_etal1,Ponty_etal2,Laval_etal,Mininni_Montgomery,Bayliss_etal} have been  
motivated by the liquid-metal dynamo experiments 
to investigate the dynamo action at low $\Pm$ in numerical 
simulations where the forcing was spatially inhomogeneous and constant 
in time rather than random. The velocity field in these simulations consisted 
of a time-independent mean flow and an energetically 
a few times weaker fluctuating component (turbulence).
The stability curves $\Rmc(\Re)$ obtained in these studies 
have an entirely different origin than ours. 
In order to illustrate the difference, \figref{fig_Rmc} shows 
the stability curves for simulations with Taylor-Green forcing, using 
published data \cite{Ponty_etal1,Ponty_etal2}. 
We see that the dynamo threshold for the simulations with a mean 
flow is much lower than for our homogeneous simulations. 
The difference is not merely quantitative.
The mean flows in question are {\em mean-field dynamos}
(even in the case of the nonhelical Taylor-Green forcing). 
This is confirmed by the ordered box-scale structure of the 
growing magnetic field reported for these simulations at $\Pm\ge1$ 
(the lower part of their stability curve). 
For $\Pm\ge1$, the threshold for the field amplification 
is $\Rmc\sim10$, which is a typical situation 
for mean-field dynamos \cite{Brandenburg}. 
The presence of magnetic energy at small scales 
is probably due to the random tangling of the mean field by turbulence, 
rather than to the fluctuation dynamo, because $\Rm$ 
is below the fluctuation-dynamo threshold. 
The increase of $\Rmc$ with $\Re$ in these simulations 
is due to the interference by the turbulence with the dynamo 
properties of the mean flow \cite{Petrelis_Fauve,Laval_etal}. 
It has not been checked whether the turbulence in 
these simulations might itself be a dynamo. 
Comparison of the two stability 
curves in \figref{fig_Rmc} suggests that this can only 
happen at much larger $\Rm$ than studied so far. 

\paragraph{Outstanding Questions.} 
The most important factual issue that remains unresolved 
is whether the low-$\Pm$ dynamo we have found is due 
to the inertial-range motions. The most important physical 
question is what is the physical mechanism that makes the dynamo
possible. 

If the local (in scale space) interaction of the inertial-range 
motions with the magnetic field is capable of amplifying the field, 
the dominant contribution to such a dynamo should be from 
the motions at the resistive scale 
$\lres\sim \lt({\eta^3/\eps}\rt)^{1/4} \sim \lf\Rm^{-3/4}$ 
\cite{Moffatt}, where the stretching rate is maximal. 
The growth rate of the magnetic energy should scale as the 
stretching rate: 
$\gamma \sim (\eps/\eta)^{1/2} \sim (\urms/\lf)\Rm^{1/2}$. 
For $\Rm\gg1$, such a dynamo would 
always be faster than a mean-field or any other kind of dynamo 
associated with the outer-scale motions, because the latter 
cannot amplify the field faster than at the rate 
$\sim \urms/\lf$. Thus, the most pressing task 
for future numerical studies is to determine whether 
$\gamma$ scales as $\Rm^{1/2}$ or reaches an $\Rm$-independent 
limit. If the latter is the case, one will have to conclude that 
it is the outer-scale motions that act 
as a dynamo despite (or in concert with) the turbulence in 
the inertial range. Unlike the inertial-range dynamo, the 
characteristics of such a dynamo would not be universal.

Theoretical predictions for a Gaussian 
white-noise velocity field (the Kazantsev \cite{Kazantsev} 
model) strengthen the case for an inertial-range dynamo. 
For a certain range of scaling exponents of 
the velocity correlation function, it is 
possible to prove that the Kazantsev field 
is a dynamo \cite{Rogachevskii_Kleeorin,Boldyrev_Cattaneo}. 
The dynamo threshold for $\Pm\ll1$ 
is predicted to be $\Rmcinf\simeq400$ (using our definition 
of $\Rm$) --- an overestimate by only (or by as much as) a factor of~2. 
The difficulty in deciding whether this theory applies 
lies in the unknown effect that assuming zero correlation 
time has on the dynamo properties of the inertial-range velocity 
field (in real turbulence, this correlation time 
is not only not small but also scale dependent). 
The main problem is that the result of Kazantsev theory is 
purely mathematical and that we do not have a physical model of the 
inertial-range dynamo.

\paragraph{Conclusions.} We have established that the  
fluctuation dynamo exists in nonhelical randomly forced 
homogeneous turbulence of a conducting fluid with low $\Pm$. 
The critical $\Rm$ in this regime 
is approximately 3 times larger than for $\Pm\ge1$. 
The nature of the dynamo and its stability curve 
$\Rmc(\Re)$ are different from the low-$\Pm$ dynamo found 
in simulations and liquid-metal experiments with a mean flow.
The physical mechanism that enables the sustained 
growth of magnetic fluctuations in the low-$\Pm$ regime 
is unknown. Is is not as yet possible to determine numerically 
whether the fluctuation dynamo is driven by the inertial-range motions 
at the resistive scale and consequently has a growth rate $\propto\Rm^{1/2}$, 
or rather is an outer-scale effect and has a constant growth rate comparable 
to the turnover rate of the outer-scale motions. 


\begin{acknowledgments}
We have benefited from discussions with 
S.~Fauve, N.~Kleeorin, J.-F.~Pinton, and I. Rogachevskii.
We thank V.~Decyk who kindly let us use his FFT libraries. 
A.B.I.\ was supported by the USDOE Center for Multiscale Plasma Dynamics, 
A.A.S.\ by PPARC. 
Simulations were done on NCSA (Illinois), UKAFF (Leicester), and 
the Dawson cluster (UCLA). 
\end{acknowledgments}

\bibliography{iscmp_PRL}
 
\end{document}